\documentclass[12pt]{article}
\usepackage[a4paper,margin=2cm]{geometry}
\usepackage[english]{babel}
\usepackage[T1]{fontenc}
\usepackage[latin1]{inputenc}
\usepackage{times,color}
\usepackage{fancyhdr,enumerate,multirow}
\usepackage{amsmath,graphicx,amssymb}

\title{\textbf{A simple hysteretic constitutive model for unsaturated flow}}

\author{Mariangeles Soldi $^{(1,*)}$, Luis Guarracino $^{(1)}$, Damien Jougnot $^{(2)}$}
		
\begin{document}
%\marginsize{2.5cm}{2cm}{2cm}{2cm} %con esto defino los márgenes, orden: izq, der, sup, inf

%\date{}
\maketitle

(1) Facultad de Ciencias Astron\'omicas y Geof\'isicas, Universidad Nacional de La Plata, Consejo Nacional de Investigaciones Cient\'ificas y T\'ecnicas, La Plata, Argentina      
              
(2) Sorbonne Universites, UPMC Univ Paris 06, CNRS, EPHE, UMR 7619 METIS, Paris, France

(*)Corresponding author: msoldi@fcaglp.unlp.edu.ar

\vskip 1cm
\textit{ \\This paper has been published in Transport in Porous Media, please cite as:\\
M. Soldi, L. Guarracino, D. Jougnot (2017) A simple hysteretic constitutive model for unsaturated flow. Transp Porous Media 120(2) 271 - 285, doi:10.1007/s11242-017-0920-2.}

\begin{abstract}
In this paper we present a constitutive model to describe unsaturated flow that considers the hysteresis phenomena. This constitutive model provides simple mathematical expressions for both saturation and hydraulic conductivity curves, and a relationship between permeability and porosity. The model is based on the assumption that the porous media can be represented by a bundle of capillary tubes with throats or ``ink-bottles'' and a fractal pore size distribution. Under these hypotheses, hysteretic curves are obtained for saturation and relative hydraulic conductivity in terms of pressure head. However, a non-hysteretic relationship is obtained when relative hydraulic conductivity is expressed as a function of saturation. The proposed relationship between permeability and porosity is similar to the well-known Kozeny-Carman equation but depends on the fractal dimension. The performance of the constitutive model is tested against different sets of experimental data and previous models. In all of the cases the proposed expressions fit fairly well the experimental data and predicts values of permeability and hydraulic conductivity better than others models.
\end{abstract}

%----------------------------------------------------------------------------------
\section{Introduction}
\label{intro}
%----------------------------------------------------------------------------------
Constitutive models for unsaturated flow provide relationships between saturation (or water content), hydraulic conductivity and pressure head. These relationships define the hydraulic behavior of soils and are necessary for the numerical resolution of the non-linear Richards equation \cite{richards1931}. From a numerical point of view, it is desirable that the mathematical expressions of the constitutive model have simple analytical forms with a small number of parameters in order to reduce the computational cost of each iteration of the linearization method. In the last decades, several empirical and semi-empirically models have been proposed, being the most widely used the van Genuchten \cite{vangenuchten1980} and the Brooks and Corey \cite{brooks1964} models. Van Genuchten proposed an empirical relation for saturation to obtain a closed-form analytical expression for the hydraulic conductivity by using Burdine \cite{burdine1953} and Mualem \cite{mualem1976} predictive models. Similarly, the Brooks and Corey model combines a power-law relation for saturation with Burdine model to obtain a simple closed-form analytical expression for the hydraulic conductivity. More recently, Assouline et al. \cite{assouline1998} proposed a conceptual model for saturation based on the assumption that the soil structure results from a uniform random fragmentation process. Then, Assouline \cite{assouline2001} developed a model to predict the relative hydraulic conductivity based on the first two moments of the water retention curve. In the particular case of fractured rocks, a physical constitutive model based on fractal geometry has been proposed by Guarracino \cite{guarracino2006} and, Monachesi and Guarracino \cite{monachesi2011}. \\
Constitutive models describe hydraulic parameters at the representative elementary volume (REV) scale. The water flow in a REV is usually described by capillary tube models with different shapes and pore size distributions. Most models assume circular cross-sectional capillary tubes, but recently Wang et al. \cite{wang2015} proposed a permeability model assuming arbitrary cross-sectional shapes of the tubes. Different approaches have been introduced to represent pore-size distributions, e.g. multi-modal, Gaussian and fractal distributions (e.g. Rubin \cite{rubin1967}, Topp \cite{topp1971}, Poulovassilis and Tzimas \cite{poulovassilis1975}, Jerauld and Salter \cite{jerauld1990}, Xu and Torres-Verd\'in \cite{xu2013}, Guarracino et al. \cite{guarracino2014}). Fractal distribution are commonly used to characterize porous media due to its simplicity and its capacity to describe a wide range of problems and soil textures (e.g. Tyler and Wheatcraft \cite{tyler1990}, Yu et al. \cite{yu2003}, Yu and Li \cite{yu2001}). In particular, Ghanbarian-Alavijeh et al. \cite{ghanbarian2011} propose a fairly recent review that illustrates the use of fractals to parameterize water retention curves.\\
Hydraulic properties of porous media present hysteresis phenomena which can significantly influence the flow and transport in partially saturated soils (e.g. Rubin \cite{rubin1967}, Topp \cite{topp1971},  Poulovassilis and Tsimas \cite{poulovassilis1975}, Jerauld and Salter \cite{jerauld1990}). Hysteresis refers to the non-unique relationships between pressure head and both saturation and hydraulic conductivity. This phenomena depends on the water movement history during the imbibition and drying processes and is mainly believed to be caused by irregularities in the cross-section of the pores or ``ink-bottle'' effects, contact angle effects or entrapped air (Jury et al. \cite{jury1991}, Klausner \cite{klausner1991}). Modeling of hysteresis requires knowledge of at least one branch of the main hysteresis loop (Mualem \cite{mualem1977}). In their review Pham et al. \cite{pham2005} divided hysteretic models into two main groups: domain models or physically based (e.g., N\'eel \cite{neel1942}, Mualem \cite{mualem1973}) and empirical models (e.g., Feng and Fredlund \cite{feng1999}, Karube and Kawai \cite{karube2001}). \\
In this study we derive a constitutive model for unsaturated flow assuming a porous media conceptualized as a bundle of constrictive capillary tubes with a fractal pore-size distribution. The tubes present pore throats or ``ink-bottles'' that allows to introduce the hysteresis in a simple form and also to characterize soils with high porosity and low permeability. Analytical closed-form expressions are obtained for saturation and hydraulic conductivity curves which are easy to evaluate and show a good agreement with experimental data. The proposed expressions have four independent physical and geometrical parameters: the fractal dimension of the pore-size distribution, a radial factor that characterize the size of the pore throat, and the maximum and minimum values of pressure head. In addition, an expression for the permeability as a function of porosity is obtained from the proposed model which becomes similar to the Kozeny-Carman equation but shows a better agreement with different experimental data sets.

%----------------------------------------------------------------------------------
\section{Constitutive model}
\label{sec:2}
%----------------------------------------------------------------------------------
In this section, we derive closed-form analytical expressions for saturation and hydraulic conductivity curves. First, we present the pore geometry of the proposed model and we derive some hydraulic properties which are valid for a single pore. Then, assuming a cylindrical REV of porous media with a fractal pore-size distribution, we obtain expressions for porosity, saturated hydraulic conductivity, and saturation and relative hydraulic conductivity curves.

%Text with citations \cite{RefB} and \cite{RefJ}.
%----------------------------------------------------------------------------------
\subsection{Hydraulic description at pore scale}
\label{sec:pore-scale}
\begin{figure}
\centering
%  \includegraphics{tubo-paper28.eps}
  \includegraphics[width=0.5\textwidth,keepaspectratio=true]{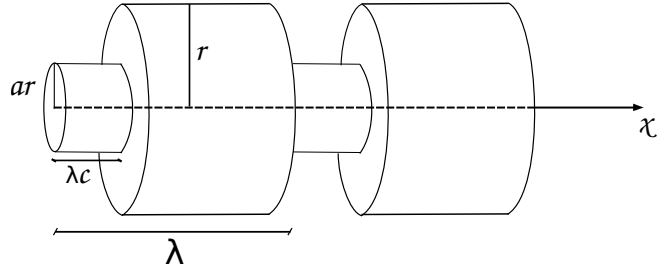}
\caption{Pore geometry of a single capillary tube with periodic pore throats}
\label{fig:modelo1}       % Give a unique label
\end{figure}

%----------------------------------------------------------------------------------
The porous media is represented by a bundle of constrictive capillary tubes. Each pore is conceptualized as a cylindrical tube of radius $r$ and length $L$ with periodically throats represented by a segment of the tube with a smaller radius, as illustrated in Figure~\ref{fig:modelo1}.

%For one-column wide figures use

Assuming that the pore geometry has a wavelength $\lambda$ and that the length of the tube $L$ contains an integer number $N$ of wavelengths, the pore radius along the tube can be described as follows:
\begin{equation}
r(x)= \left\{ \begin{array}{lll}
ar   \quad & \text{if} \quad  & x\in  [0 + 2\pi n  , \lambda c + 2\pi n )\\
r    \quad & \text{if} \quad  & x\in  [\lambda c + 2\pi n ,\lambda + 2\pi n),
\end{array} \right.
\label{eq:radio}
\end{equation}
where $a$ is the \emph{radial factor} ($0\leq a \leq 1$), $c$ is the \emph{length factor} of the pore throat ($0\leq c \leq 1$) and $n = 0,1,...,N-1$. The parameter $a$ represents the ratio in which the radius is reduced, and the parameter $c$ represents the fraction of $\lambda$ with a narrow neck. Note that if $c = 1$ or $c = 0$ we obtain a straight tube with radii $ar$ or $r$, respectively.

\begin{figure}
\centering
\includegraphics[width=0.9\textwidth,keepaspectratio=true]{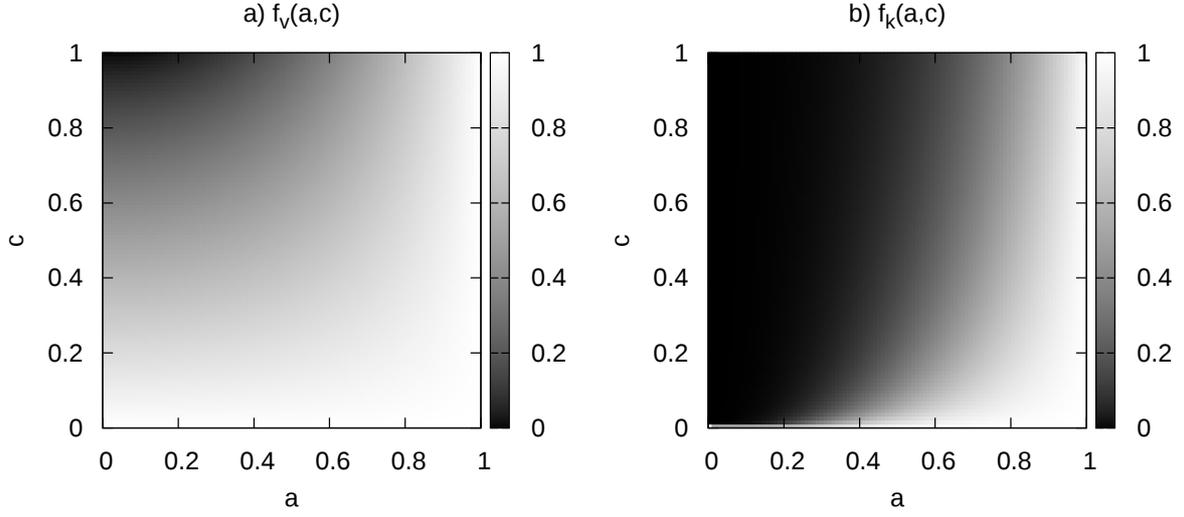}
\caption{Dimensionless factors $f_v(a,c)$ and $f_k(a,c)$. These factors control the pore volume and the volumetric flow rate at pore scale, and the porosity and saturated hydraulic conductivity at REV scale.}
\label{fig:density}
\end{figure}

Based on the above assumptions, the volume of a single tube can be calculated by integrating the crossectional area over the length $L$ as follows:
\begin{equation}
V_p (r) = \int_0^L \pi r^2(x) dx = N \left[ \int_0^{\lambda c} \pi r^2 dx + \int_{\lambda c}^{\lambda} \pi (ra)^2 dx \right] = L \pi r^2 f_v (a,c),
\label{eq:vol-poro}
\end{equation}
where 
\begin{equation}
f_v (a,c)= a^2 c +1-c ,
\label{eq:f_v}
\end{equation}
$f_v$ is a factor that varies between $0$ and $1$, and quantifies the reduction of pore volume due to the constrictivity of the tube. A density plot of $f_v$ is shown in Figure \ref{fig:density}(a). Note that low values of parameter $c$ or large values of parameter $a$ produce small variations of the pore volume.

Under the assumption of laminar flow and ignoring the convergence and divergence of the flow, the volumetric flow rate of a pore with periodical varying aperture $Q_p(r)$ can be approximated with (Bodurtha \cite{bodurtha2003}, Bousfield and Karles \cite{bousfield2003}):
\begin{equation}
Q_p (r) = \frac{\rho g}{\mu} \left[ \frac{1}{L} \int_0^L \frac{8}{\pi r^4 (x)} dx \right]^{-1} \frac{\Delta h}{L},
\label{eq:Q-poro}
\end{equation}
where $\rho$ is the water density, $g$ gravity, $\mu$ water viscosity and $\Delta h$ the head drop across the tube.

Substituting Eq. \eqref{eq:radio} in Eq. \eqref{eq:Q-poro} yields:
\begin{equation}
Q_p(r) = \frac{\rho g}{\mu} \frac{\pi r^4}{8} f_k(a,c) \frac{\Delta h}{L},
\label{eq:Q_p}
\end{equation}
where
\begin{equation}
f_k(a,c) = \frac{a^4}{c + a^4(1 - c)},
\label{eq:f_k}
\end{equation}
$f_k$ is a factor that quantifies the volumetric flow rate reduction due to pore throats and varies between 0 and 1. Figure~\ref{fig:density}(b) shows the variation of $f_k$  as a function of the radial factor $a$ and the length factor $c$. As it can be expected low values of $a$ drastically reduce the volumetric flow rate of the tube.

If we compare Figures \ref{fig:density}(a) and \ref{fig:density}(b), it can be noticed that the values of the parameters $a$ and $c$ modify the volume and the volumetric flow rate of the tube in different ways. For example, for low values of $a$ and $c$, the volume of the pore is slightly affected while the volumetric flow of the pore significantly decreases. Also note that for $a = 1$ or $c = 0$, $f_v = f_k = 1$, and the expressions obtained for Eqs. \eqref{eq:vol-poro} and \eqref{eq:Q_p} represent the volume and the volumetric flow rate of a straight tube of radius $r$, respectively.

%------------------------------------------------------------------------------------------------
\subsection{Hydraulic description at REV scale}
\label{sec:rev}
%------------------------------------------------------------------------------------------------
To derive the expressions for saturation and hydraulic conductivity, we consider as a REV a straight circular cylinder of radius $R$ and length $L$. The choice of the REV geometry is based on the shape of soil samples commonly used in laboratory tests. Other geometries, like rectangular REV, can also be considered by introducing minor changes in model derivation. The pore structure of the REV is represented by a bundle of constrictive tubes (as described in the previous section) with a fractal pore-size distribution. We also assume that the pore radius $r$ varies from a minimum value $r_{min}$ to a maximum value $r_{max}$. 

The cumulative size distribution of pores is assumed to obey the following fractal law (e.g. Tyler and Wheatcraft \cite{tyler1990}, Yu \cite{yu2008}, Guarracino \cite{guarracino2007}):
\begin{equation}
N(r) = \left(\frac{r}{R}\right)^{-D},
\label{eq:cantporos}
\end{equation}
where $D$ ($1<D<2$) is the fractal dimension. Note that if $r_{max}=R$, $N = 1$ and the REV is fully occuped by a single pore. On the other hand, if $r_{min}=0$, the REV contains an infinite number of pores. 

Differentiating Eq. \eqref{eq:cantporos} with respect to $r$ we obtain the number of pores whose sizes are within the infinitesimal range $r$ and $r+dr$:
\begin{equation}
dN(r) = -D R^D r^{-D-1}dr.
\label{eq:fractal}
\end{equation}
The negative sign in Eq. \eqref{eq:fractal} implies that the pore number decreases with the increase of the pore size (Yu \cite{yu2003}).

The porosity $\phi$ of the REV, can be computed from its definition:
\begin{equation}
\phi = \frac{\text{Volume of pore space}}{\text{Volume of REV}} = \frac{\int_{r_{min}}^{r_{max}} V_p(r) dN(r)}{\pi R^2 L}.
\label{eq:fi}
\end{equation}
Replacing Eqs. \eqref{eq:vol-poro} and \eqref{eq:fractal} into Eq. \eqref{eq:fi}, the porosity of the REV can be expressed as:
\begin{equation}
\phi = f_v \phi^{ST},
\label{eq:poroST}
\end{equation}
where
\begin{equation}
\phi^{ST} = \frac{D}{R^{(2-D)}(2-D)} \left[ r_{max}^{2-D} - r_{min}^{2-D} \right]
\label{eq:porosidad}
\end{equation}
is the porosity of the REV considering straight tubes (i.e., $a=1$).

The volumetric flow rate $Q$ at REV scale can be obtained by integrating all the pores volumetric flow rates given by Eq. \eqref{eq:Q-poro} over the entire range of pore sizes:
\begin{equation}
Q = \int_{r_{min}}^{r_{max}} Q_p(r) dN(r) = 
 \frac{\rho g}{\mu} \frac{f_k}{8} \frac{\Delta h}{L} \frac{\pi D R^D}{(4-D)} \left[r_{max}^{4-D}-r_{min}^{4-D}\right].
 \label{eq:caudal}
\end{equation}

On the other hand, on the basis of Darcy's law \cite{darcy1856}, the volumetric flow rate through the REV can be expressed as:
\begin{equation}
Q = K_s \frac{\Delta h}{L} \pi R^2,
\label{eq:Darcy}
\end{equation}
where $K_s$ is the saturated hydraulic conductivity. Combining Eqs. \eqref{eq:caudal} and \eqref{eq:Darcy} an expression for $K_s$ yields:
\begin{equation}
K_s = f_k K_s^{ST},
\label{eq:KST}
\end{equation}
where
\begin{equation}
K_s^{ST} = \frac{\rho g}{\mu} \frac{1}{8} \frac{D}{R^{(2-D)}(4-D)} \left[r_{max}^{4-D}-r_{min}^{4-D}\right]
\label{eq:Ksat}
\end{equation}
is the saturated hydraulic conductivity of the REV considering straight tubes.

By inspection of Eqs. \eqref{eq:poroST} and \eqref{eq:KST}, it can be noticed that the factors $f_v$ and $f_k$ produce different changes in the macroscale properties of the REV $\phi$ and $K_s$, respectively. It can be demonstrated that for every value of parameters $a$ and $c$, $f_k$ is always smaller than $f_v$ allowing us to represent media with high porosity, low permeability and low specific surface area. Our model is also able to describe media which have the same porosity but different permeabilities. For example, clay and sand soils have typically similar porosities but their hydraulic conductivities differs by several orders of magnitude (e.g. Carsel and Parrish \cite{carsel1988}). 

For most porous media $r_{min}/r_{max} \simeq 10^{-2}$ (Yu and Li \cite{yu2001}), then we can assume that $r_{min} \ll r_{max}$. Under the above assumption, the terms $r_{min}^{2-D}$ and $r_{min}^{4-D}$ in Eqs. \eqref{eq:porosidad} and \eqref{eq:Ksat} can be considered negligible. Then, combining the resulting expressions, we can obtain the following relationship between $K_s$ and $\phi$:

\begin{equation}
K_s = \alpha f_k \left[\frac{\phi}{f_v}\right]^{\left(\frac{4-D}{2-D}\right)},
\label{eq:aproxKC}
\end{equation}
where
\begin{equation}
\alpha = \frac{\rho g}{\mu} \frac{D R^2}{8(4-D)} \left[\frac{2-D}{D}\right]^{\left(\frac{4-D}{2-D}\right)}.
\label{eq:alpha}
\end{equation}

Note that the exponent of porosity in Eq. \eqref{eq:aproxKC} $(4-D)/(2-D)$ is greater than 3. In the limit case of a cubic exponent Eq. \eqref{eq:aproxKC} becomes similar to the KC equation. This issue will be further analyzed in Section \ref{sec:kozeny} where Eq. \eqref{eq:aproxKC} is tested against experimental data sets.

\subsection{Saturation and relative hydraulic conductivity curves}
\label{sec:curves}
%------------------------------------------------------------------------------------------------
In this section, we derive the saturation and relative hydraulic conductivity curves of the constitutive model. Due to the varying aperture of the pores, the retention curves obtained from drainage and imbibition tests are expected to be different. The hysteresis phenomena can be easily introduced in the model thanks to the pore geometry illustrated in Fig. \ref{fig:modelo1} and described by Eq. \eqref{eq:radio}. 

For a straight tube, we can relate the radius of the water-filled pore $r_h$ to the pressure head $h$ by the following equation (Jurin \cite{jurin1717}, Bear \cite{bear1998}):
\begin{equation}
h = \frac{2 \sigma \cos (\beta)}{\rho g r_h},
\label{eq:h}
\end{equation}
where $\sigma$ is the surface tension of the water and $\beta$ the contact angle.

To obtain the main drying saturation curve, we consider that the REV is initially fully saturated and is drained by a pressure head $h$. We assume that a tube becomes fully desaturated if the radius of the pore throat $ar$ is greater than the radius  $r_h$ given by Eq. \eqref{eq:h}. Then it is reasonable to also assume that pores with radii $r$ between $r_{min}$ and $r_h/a$ will remain fully saturated. Therefore, according to Eqs. \eqref{eq:vol-poro} and \eqref{eq:fractal}, the drying saturation curve $S^d_{e}$ can be computed by:
\begin{equation}
S^d_{e} = \frac{\int_{r_{min}}^{\frac{r_h}{a}} V_p(r) dN(r)}{\int_{r_{min}}^{r_{max}} V_p(r) dN(r)} = \frac{\left(\frac{r_h}{a}\right)^{2-D} - r_{min}^{2-D}}{r_{max}^{2-D} -r_{min}^{2-D} }.
\label{eq:sd1}
\end{equation}

Substituting Eq. \eqref{eq:h} into Eq. \eqref{eq:sd1} yields

\begin{equation}
S^d_{e}(h)= \left\lbrace 
\begin{array}{ll}
1 \quad & \text{if} \quad h \leq \frac{h_{min}}{a} \\ \\
\frac{\left(ha\right)^{D-2} - h_{max}^{D-2}}{h_{min}^{D-2} -h_{max}^{D-2} } \quad & \text{if} \quad \frac{h_{min}}{a} \leq h \leq \frac{h_{max}}{a}, \\ \\
0 \quad & \text{if} \quad h \geq \frac{h_{max}}{a} \\
\end{array} \right.
\label{eq:sd}
\end{equation}
where
\begin{equation}
h_{min} = \frac{2 \sigma \cos (\beta)}{\rho g r_{max}} \qquad h_{max} = \frac{2 \sigma \cos (\beta)}{\rho g r_{min}},
\label{eq:h_min-max}
\end{equation}
$h_{min}$ and $h_{max}$ being the minimum and maximum pressure heads defined by $r_{max}$ and $r_{min}$, respectively.

Similarly, the main wetting saturation curve can be obtained assuming that the REV is initially dry and it is flooded with a pressure $h$. In this case, only the tubes with radius $r$ smaller than $r_h$ will be fully saturated. Then the main wetting saturation curve $S^w_{e}$ can be expressed as:

\begin{equation}
S^w_{e}(h)= \left\lbrace
\begin{array}{ll}
1 \quad & \text{if} \quad h \leq h_{min} \\ \\
\frac{h^{D-2} - h_{max}^{D-2}}{h_{min}^{D-2} -h_{max}^{D-2} } \quad & \text{if} \quad h_{min} \leq h \leq {h_{max}}. \\ \\
0 \quad & \text{if} \quad h \geq h_{max} \\
\end{array} \right.
\label{eq:sw}
\end{equation}

Using the same hypothesis and neglecting film flow on tube surfaces, we can obtain the main drying and wetting curves for relative hydraulic conductivity. During drainage only tubes with pore throat radius $ar$ smaller than $r_h$ are fully saturated. Then, the main contribution to the total volumetric flow through the REV can be obtained by integrating the individual volumetric flow rates $Q_p$ given by Eq. \eqref{eq:Q_p} over the pores that remain fully saturated ($r_{min} \leq r \leq r_h/a$):
\begin{equation}
Q = \int_{r_{min}}^{\frac{r_h}{a}} Q_p(r) dN(r) .
\label{eq:Qunsat}
\end{equation}

Otherwise, according to Buckingham-Darcy's law \cite{buckingham1907}, the total volumetric flow rate through the REV can be expressed as:
\begin{equation}
Q = K_s K_r(h) \frac{\Delta h}{L} \pi R^2,
\label{eq:B-Darcy}
\end{equation}
where $K_r$ is the relative hydraulic conductivity which is a dimensionless function of $h$ and varies between $0$ and $1$.
 
Combining Eqs. \eqref{eq:Qunsat} and \eqref{eq:B-Darcy}, and using Eqs. \eqref{eq:Q_p}, \eqref{eq:fractal} and \eqref{eq:KST} we obtain the relative hydraulic conductivity for the drying process:
\begin{equation}
K^d_r = \frac{\left(\frac{r_h}{a}\right)^{4-D} - r_{min}^{4-D}}{r_{max}^{4-D} -r_{min}^{4-D} }
\label{eq:kd1}
\end{equation}

Finally, using Eq. \eqref{eq:h} we can express Eq. \eqref{eq:kd1} in terms of pressure head:

\begin{equation}
K^d_r(h)= \left\lbrace
\begin{array}{ll}
1 \quad & \text{if} \quad h \leq \frac{h_{min}}{a} \\ \\
\frac{\left(ha \right)^{D-4} - h_{max}^{D-4}}{h_{min}^{D-4} - h_{max}^{D-4} } \quad & \text{if} \quad \frac{h_{min}}{a} \leq h \leq \frac{h_{max}}{a}. \\ \\
0 \quad & \text{if} \quad h \geq \frac{h_{max}}{a} \\
\end{array} \right.
\label{eq:kd}
\end{equation}

Similarly, the main wetting relative hydraulic conductivity curve $K^w_r(h)$ can be derived by integrating Eq. \eqref{eq:Qunsat} over the range of saturated pores ($r_{min} \leq r \leq r_h$):
\begin{equation}
K^w_r(h)= \left\lbrace
\begin{array}{ll}
1 \quad & \text{if} \quad h \leq h_{min} \\ \\
\frac{h^{D-4} - h_{max}^{D-4}}{h_{min}^{D-4} - h_{max}^{D-4} } \quad & \text{if} \quad h_{min} \leq h \leq h_{max}. \\ \\
0 \quad & \text{if} \quad h \geq h_{max} \\
\end{array} \right.
\label{eq:kw}
\end{equation}

Note that saturation and relative hydraulic conductivity expressions for both drying and wetting have analytical closed-forms with only four independent parameters with geometrical and physical meaning: $a, D, h_{min}$ and $h_{max}$.

In the classical models of hysteresis, saturation and relative hydraulic conductivity values are limited by main drying and wetting curves which are obtained for initially fully saturated and dry porous media, respectively. For any intermediate state that does not correspond to a fully saturated or dry medium, scanning curves can be scaled from the main drying and wetting curves for both relationships, $S_e(h)$ and $K_r(h)$. These scanning curves can be generated using different approaches like play-type (Beliaev and Hassanizadeh \cite{beliaev2001}) or scaling hysteresis (Parker and Lenhard \cite{parker1987}).

Relative hydraulic conductivity $K_r$ can also be expressed in terms of saturation $S_{e}$. By combining Eqs. \eqref{eq:sd1} and \eqref{eq:kd1} we obtain the following unique equation for the drying and the wetting:

\begin{equation}
K_{r} = \frac{{\left\lbrace S_{e} \left[ \left( \frac{h_{min}}{h_{max}}\right)^{D-2} - 1 \right] +1 \right\rbrace}^{\frac{D-4}{D-2}} -1}{\left( \frac{h_{min}}{h_{max}}\right)^{D-4} - 1}.
\label{eq:KdeS}
\end{equation}
It is interesting to remark that the relationship $K_r(S_{e})$ results in a nonhysteretic function across the entire range of saturations and this result is in agreement with a number of experimental data  (e.g. van Genuchten \cite{vangenuchten1980}, Mualem \cite{mualem1986}, Topp and Miller \cite{topp1966}).

For $h_{max} \gg h_{min}$, Eq. \eqref{eq:KdeS} can be reduced to: 
\begin{equation}
K_r = {S_{e}}^{\frac{D-4}{D-2}},
\label{eq:K(S)}
\end{equation}
which is consistent with the well-known Brooks and Corey model, $K_r = {S}^{\frac{2+3\lambda}{\lambda}}$, where $\lambda$ is a dimensionless and empiric parameter related to the pore size distribution. Parameter $\lambda$ can be related with the fractal dimension $D$ through $\lambda = (D-2)/(1-D)$. Considering the range of $\lambda$ values between 0.21 and 3.02 reported by Assouline \cite{assouline2005} for different type of soils, it yields values of $D$ between $1.249$ and $1.826$ which are consistent with the admissible range of $D$ values.

%------------------------------------------------------------------------------------------------------------
\section{Comparison with experimental data}
\label{sec:exp-data}
%------------------------------------------------------------------------------------------------------------
In the present section, we test the ability of the proposed model to reproduce available measured data from the research literature. These data sets consists on measured permeability-porosity, relative hydraulic conductivity-saturation, and hysteretic saturation-pressure head values for different soil textures. 

%------------------------------------------------------------------------------------------------------------
\subsection{Permeability}
\label{sec:kozeny}
%------------------------------------------------------------------------------------------------------------
In order to test the proposed relationship between permeability and porosity for different types of soils, we selected four data series from Luffel et al. \cite{luffel1991}, Hirst et al. \cite{hirst2001} and Chilindar \cite{chilindar1964}. As it is well known, permeability $k$ and saturated hydraulic conductivity $K_s$ are related through $K_s = k \rho g / \mu$. According to Eq. \eqref{eq:aproxKC}, the proposed permeability model can be expressed as follows: 
\begin{equation}
k = C \phi^{\left(\frac{4-D}{2-D}\right)},
\label{eq:KCproposed}
\end{equation}
where
\begin{equation}
C = \frac{\mu}{\rho g} \alpha f_k f_v^{\left(\frac{2-D}{4-D}\right)}.
\label{eq:coef-KC-model}
\end{equation}

Eq. \eqref{eq:KCproposed} will be also compared with the Kozeny-Carman equation which reads (Kozeny \cite{kozeny1927}, Carman \cite{carman1937}):
\begin{equation}
k = C_{KC} \frac{\phi^3}{(1-\phi)^2}
\label{eq:KC}
\end{equation}
where $C_{KC}$ is a parameter that depends on the specific internal surface area, an empirical geometrical parameter and the tortuosity.

\begin{figure}
\centering
\includegraphics[width = 4.8in]{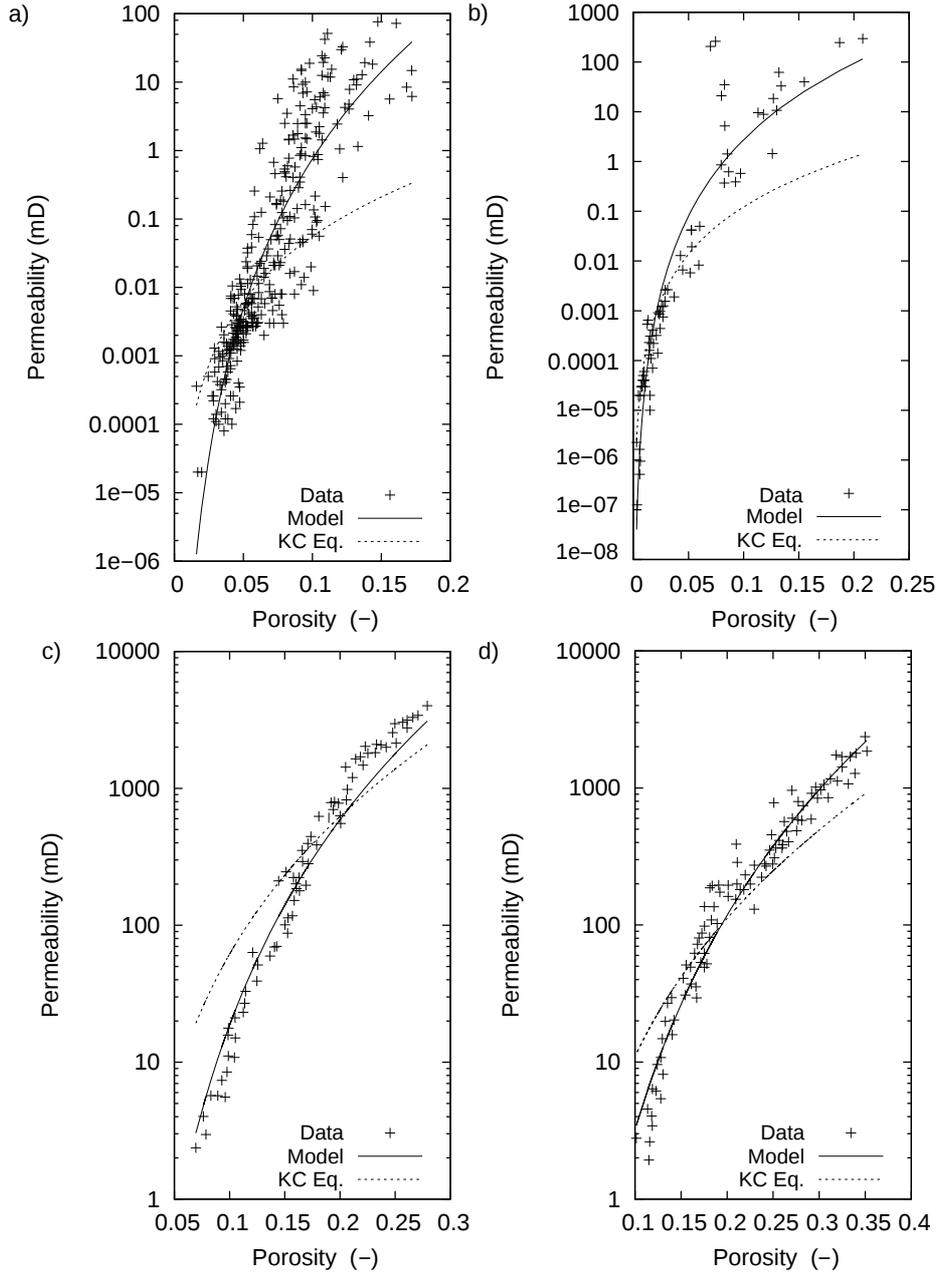}
\caption{Comparison between the proposed model, the KC equation and experimental data sets of permeability-porosity. (a) Early Cretaceous Fluvial and Deltaic Channel Sandstones (data from Luffel et al. \cite{luffel1991}), (b) Carboniferous and Devonian Timimoun Basin ("tight gas" sandstones)(data from Hirst et al. \cite{hirst2001}), (c) Fined grained sandstone and (d) silty sandstones (data from Chilindar \cite{chilindar1964}).}
\label{fig:curvasKC}
\end{figure}

For each type of soil Eqs. \eqref{eq:KCproposed} and \eqref{eq:KC} are fitted to measured data by minimizing the root-mean-square deviation (RMSD):
\begin{equation}
RMSD = \left\lbrace \frac{1}{n} \sum_{i=1}^{n} \left[ \log (k_i^{calc}) - \log (k_i^{dat}) \right]^2 \right\rbrace ^{0.5}
\label{eq:Rms-log}
\end{equation}
where $k^{calc}$ and $k^{dat}$ correspond to the calculated and measured permeabilities, respectively. A logarithmic scale was considered because of the wide range of variation for the permeability values. The fitted parameters for Eqs. \eqref{eq:KCproposed} and \eqref{eq:KC} are listed in Table \ref{table:KC} as well as their respective RMSD. It can be noted that, for all soils, the RMSD of the proposed model is smaller than the ones from the KC equation. Figure \ref{fig:curvasKC} shows that the proposed relationship predicts fairly good the observed values over a range of 4 to 10 orders of magnitude.

\begin{table}%[!h]
\centering
\caption{Values of the fitted parameters ($D$, $C$ and $C_{KC}$) and the RMSD corresponding to the proposed model (Eq. \eqref{eq:KCproposed}) and to the KC equation (Eq. \eqref{eq:KC})}
\label{table:KC}
\begin{tabular}{llllll}
\hline\noalign{\smallskip}
Soil type & $D$ & $C$ (mD)& RMSD & $C_{KC}$ (mD)& RMSD$_{KC}$\\
\noalign{\smallskip}\hline\noalign{\smallskip}
Fluvial and deltaic & 1.68 & 1.336$\times$10$^{\text{7}}$ & 1.1386 & 44.85 & 1.3856\\
Timimoun Basin & 1.512 & 3.452$\times$10$^{\text{5}}$ & 1.1894 & 75 & 1.3942\\
Fined grained sandstone & 1.498 & 1.797$\times$10$^{\text{6}}$ & 0.5988 & 4.4$\times$10$^{\text{3}}$ & 0.8910\\
Silty sandstones & 1.524 & 5.1$\times$10$^{\text{5}}$ & 0.5478 & 8.95$\times$10$^{\text{3}}$ & 0.7680\\
\noalign{\smallskip}\hline
\end{tabular}
\end{table}

%------------------------------------------------------------------------------------------------------------
\subsection{Relative hydraulic conductivity}
\label{sec:K(S)}
%------------------------------------------------------------------------------------------------------------
The proposed relative hydraulic conductivity model (Eq. \eqref{eq:K(S)}) is tested against 8 experimental data series from the Mualem catalogue \cite{mualem1974} (see Table \ref{table:assouline}). These data series have been also used by Assouline to test his model which predicts $K_r$ from the first two moments of the water retention curve \cite{assouline2001}.
For each soil type, the proposed model is fitted to the measured data by minimizing the RMSD. 

\begin{figure}
\centering
\includegraphics[width = 4.8in]{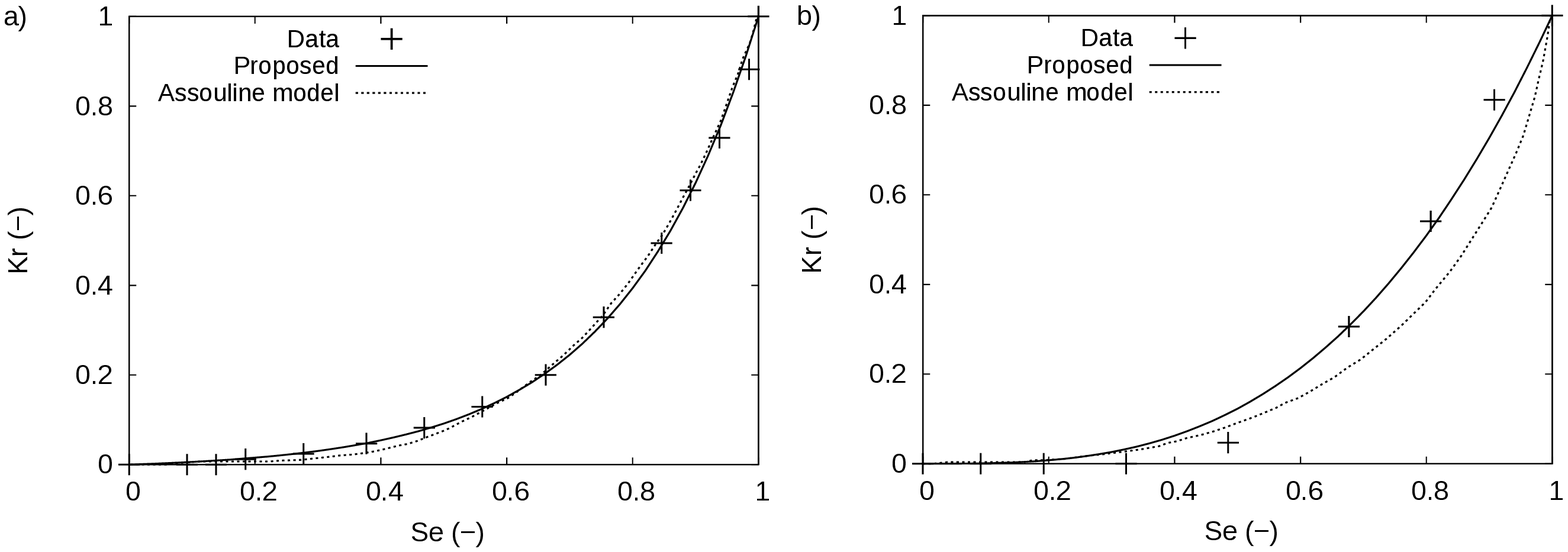}
\caption{Comparison between the relative hydraulic conductivity (Eq.~\ref{eq:K(S)}) and measured data: (a) Sable de Riviere and (b) Gilat sandy loam (data from Mualem Catalogue \cite{mualem1974}). The figure also includes the fit of Assouline model \cite{assouline2001}}
\label{fig:val-KvsS}
\end{figure}

Figure \ref{fig:val-KvsS} illustrates the fit of Eq. \eqref{eq:K(S)} and Assouline model to 2 sets of experimental data (Sable de Riviere and Gilat sandy loam) using the parameters given in Table \ref{table:assouline}. It can be noticed that the proposed model shows a significant improvement over the one of Assouline for the Gilat sandy loam (see Figure \ref{fig:val-KvsS}(b)). Table \ref{table:assouline} lists the resulting best fitted parameters for the 8 experimental data series, their RMSD and the corresponding RMSD obtained by Assouline \cite{assouline2001}. For all soil types, the RMSD values of our model are smaller than the ones obtained with the Assouline model. Note that for Gilat sandy loam and Guelph loam soils, the RMSD are even 1 order of magnitude smaller.

\begin{table}%[!h]
\centering
\caption{Values of the fitted parameters ($D$ and $h_{min}/h_{max}$), the corresponding RMSD and the RMSD of Assouline's model \cite{assouline2001}}
\label{table:assouline}
\begin{tabular}{lllll}
%\centering
\hline\noalign{\smallskip}
%Soil type & $D$ & $\frac{h_{min}}{h_{max}}$ & RMSD & RMSD \par (Assouline) \\
Soil type & $D$ & $\frac{h_{min}}{h_{max}}$ & RMSD & RMSD \\
& & & & (Assouline) \\
\noalign{\smallskip}\hline\noalign{\smallskip}
Sable de riviere & 1.99 & 0.101 & 0.015 & 0.036\\
Gilat sandy loam & 1.012 & 1.09$\times$10$^{\text{-4}}$ & 0.033 & 0.252 \\
Pouder river sand & 1.112 & 1.09$\times$10$^{\text{-4}}$ & 0.071 & 0.076\\
Amarillo silty clay loam & 1.387 & 0.001 & 0.009 & 0.014\\
Rubicon sandy loam & 1.999 & 0.088 & 0.021 & 0.046\\
Guelph loam & 1.918 & 0.021 & 0.004 & 0.037\\
Weld silty clay loam & 1.508 & 0.061 & 0.036 & 0.038\\
Silt Mont Cerris soils & 1.376 & 1.09$\times$10$^{\text{-4}}$ & 0.082 & 0.188\\
\noalign{\smallskip}\hline
\end{tabular}
\end{table}

%------------------------------------------------------------------------------------------------------------
\subsection{Saturation curve hysteresis}
\label{sec:hysteretic}
\begin{figure}
\centering
\includegraphics[width = 4.8in]{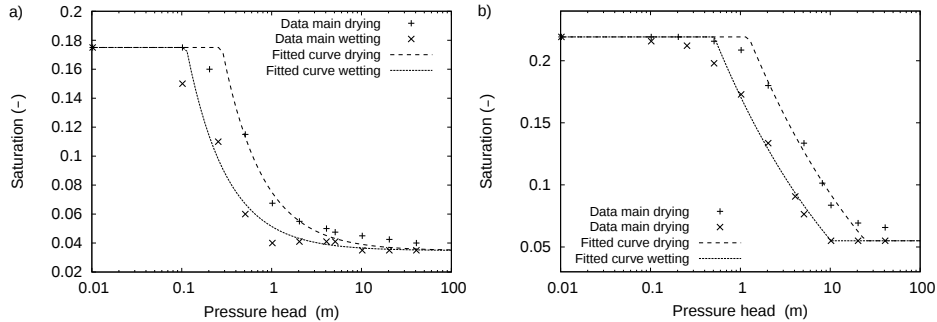}   
\caption{Comparison of the main drying and wetting saturation curves from the proposed model with experimental data sets: (a) Beaver Creek sand and (b) Processed silt (data from Pham et al. \cite{pham2003})}
\label{fig:histeresis}
\end{figure}

%------------------------------------------------------------------------------------------------------------
To test the ability of the model to describe the hysteresis phenomena we compare the main wetting and drying curves (Eqs. \eqref{eq:sd} and \eqref{eq:sw}) to experimental data from the literature. Two different soil types from Pham et al. \cite{pham2003} are used: Beaver Creek sand and a processed silt. The maximum and minimum values of pressure head (Eq. \eqref{eq:h_min-max}) were determined by trial and error method (see Table \ref{table:histeresis}). Then, the fractal dimension $D$ and the radial factor $a$ have been estimated by minimizing the RMSD between calculated and measured values of both drying and wetting curves using an exhaustive search method.
Table \ref{table:histeresis} shows the model parameters and the RMSD values for each soil. Note that even if the simplicity of the model, the hysteretic behavior of saturation can be fairly fitted by a minimum number of parameters. It is important to remark that only one set of parameters $a$ and $D$ explains both drying and wetting curves simultaneously (see Figure \ref{fig:histeresis}).

Note that the expression of the drying curve (Eq. \eqref{eq:kd}) depends on parameters $a$ and $D$. This enables us to fit these parameters using only experimental data from the main drying hysteresis loop and then to predict the main wetting curve by using Eq. \eqref{eq:kw}. Following this alternative fitting procedure, the parameter values of the drying curve are:  $a = 0.627$ and $D = 1.314$ (RMSD $= 1.877\times10^{\text{-2}}$) for the Beaver Creek sand and, $a = 0.401$ and $D = 1.722$ (RMSD $= 1.362\times10^{\text{-2}}$) for the Processed silt. Note that only the parameters of the Processed silt are similar to the ones listed in Table \ref{table:histeresis}. The prediction of the wetting curve from the drying curve could be an additional advantage of the proposed model that needs to be verified with a more exhaustive analysis and additional experimental data.

\begin{table}%[!h]
\centering
\caption{Values of the fitted parameters ($D$ and $a$) and the corresponding RMSD. Parameters $h_{min}$ and $h_{max}$ have been fixed before the estimation of $D$ and $a$}
\label{table:histeresis}
\begin{tabular}{llllll}
\hline\noalign{\smallskip}
Soil type & $D$ & $a$ & $h_{min}$ (m) & $h_{max}$ (m) & RMSD  \\
\noalign{\smallskip}\hline\noalign{\smallskip}
Beaver Creek sand & 1.0266 & 0.4008 & 0.112 & 100.0 &  1.2566$\times$10$^{\text{-2}}$ \\
Processed silt & 1.7598 & 0.4126 & 0.510 & 10.20 & 1.1178$\times$10$^{\text{-2}}$\\
\noalign{\smallskip}\hline
\end{tabular}
\end{table}

%------------------------------------------------------------------------------------------------------------
\section{Discussion and conclusion}
\label{discuss}
%------------------------------------------------------------------------------------------------------------
A physically based theoretical model for estimating the hydraulic properties for unsaturated flow in porous media has been presented. The derivation of the model relies on the assumption that porous media can be represented by a bundle of cylindrical tubes with periodically throats and a fractal pore-size distribution. Based on geometrical properties and physical laws, analytical closed-form expressions were obtained for the saturation and the relative hydraulic conductivity as functions of pressure head. These expressions contain four independent parameters ($a$, $D$, $r_{min}$ and $r_{max}$), all of them with a specific physical or geometrical meaning. It is worth mentioning that the direct determination of these parameters is a difficult task due to the need to know in detail the microscopic geometry of the porous media. Considering the current developments in imaging technology, direct measurements of the pore structure can be obtained using X-ray tomography (Wildenschild et al., \cite{wildenschild2002}). Lindquist et al. \cite{lindquist2000} applied this technique to measure distributions of channel length, throat size and pore volume of Fontainebleau sandstones.

Hysteresis in the saturation and relative hydraulic conductivity curves have been easily introduced in the model by assuming periodic constrictivities through the radial factor $a$ and the length factor $c$ (Fig. \ref{fig:modelo1}). It is interesting to note that when the relative hydraulic conductivity is expressed in terms of saturation a unique non-hysteretic relationship is obtained for both drainage and imbibition tests (Eq. \eqref{eq:K(S)}). This behavior is consistent with previous studies and experimental data  (Fig. \ref{fig:val-KvsS}, Topp and Miller\cite{topp1966}, van Genuchten \cite{vangenuchten1980}, Mualem \cite{mualem1986}). Several causes have been proposed to justify hysterersis phenomena (e.g. Jury et al. \cite{jury1991}, Klausner \cite{klausner1991}). These results enhance the hypothesis that hysteresis originates from pore throats or ``ink-bottle'' effects. Nevertheless, other effects could also explain or contribute to hysteresis in porous media, such as: network effects, contact angle hysteresis and film flow (e.g. Blunt et al. \cite{blunt2002}, Spiteri et al. \cite{spiteri2008}, Maineult et al. \cite{maineult2017}). Note that when the radial factor $a=1$ (straight tubes), the hysteresis disappears from the saturation and relative hydraulic conductivity cuves (see Sec. \ref{sec:curves}).

The presence of throats in the capillary tubes also modifies the porosity and permeability through the factors $f_v$ and $f_k$ (Eqs. \eqref{eq:poroST} and \eqref{eq:KST}), respectively. Both factors depends on $a$ and $c$, and varies between 0 and 1 (Eqs. \eqref{eq:f_v} and \eqref{eq:f_k}). Nevertheless, the factor $f_k$ that modifies the permeability is always smaller than the factor $f_v$ that affects the porosity. This allows the model to describe media with high porosity, low permeability and low specific surface area, which can not be properly represented with straight tube models.

The fractal dimension $D$ is a geometrical parameter that determines the pore-size distribution of the model.  This fractal distribution has been found to be useful to describe groundwater flow in the literature (e.g. Tyler and Wheatcraft \cite{tyler1990}, Yu et al. \cite{yu2003}, Yu and Li \cite{yu2001}, Guarracino et al. \cite{guarracino2014}, Wang et al. \cite{wang2015}). The fractal dimension can be related to the pore size distribution index $\lambda$ proposed in the Brooks and Corey model (see Sec. \ref{sec:curves}), providing a geometrical meaning to this empirical parameter.

The proposed model also provides a relationship between permeability and porosity (Eq. \eqref{eq:KCproposed}), which under simplifying assumptions is similar to the KC equation. However, the proposed model performs better than the KC equation when compared to experimental permeability data ranging over 4 to 10 orders of magnitudes (Fig. \ref{fig:curvasKC}). 

This study allowed the development of a framework to describe saturation and relative hydraulic conductivity curves that include hysteresis phenomena. The relative hydraulic conductivity has been validated using experimental data from different type of soils, showing better agreements than Assouline model (Fig. \ref{fig:val-KvsS}). The hysteretic saturation curves have also been succesfully tested with experimental data by fitting only 2 model parameters, $a$ and $D$ (Fig. \ref{fig:histeresis}).

From a mathematical point of view, all the expressions have analytical closed-forms, which are simple and easy to evaluate. Therefore, their implementation in numerical flow codes is straightforward and involves little additional computational effort compared to non-hysteretic simulations.

This simple constitutive model can be a starting point to describe other physical phenomena that require hydraulic description at pore scale, such as: generation of streaming potential (e.g. Jougnot et al. \cite{jougnot2012}), ionic transport and mixing in capillaries (e.g. Dentz et al. \cite{dentz2011}), geochemical reactions in porous media (e.g. Guarracino et al. \cite{guarracino2014}), and wave-induced fluid flow (e.g. Rubino et al. \cite{rubino2013}).

%\nolinenumbers
%EJEMPLO------------------------------------------------------
%as required. Don't forget to give each section
%and subsection a unique label (see Sect.~\ref{sec:1}).
%\paragraph{Paragraph headings} Use paragraph headings as needed.

%\begin{figure}
%  \includegraphics{example.eps}
%\end{figure}
%\begin{figure*}
%  \includegraphics[width=0.75\textwidth]{example.eps}
%\end{figure*}
%\begin{table}
%\begin{tabular}{lll}
%\hline\noalign{\smallskip}
%\end{tabular}
%\end{table}

%----------------------------------------------------------------------------------

\end{document}